\documentclass[showkeys]{revtex4}

\usepackage{epsfig}

\usepackage{amssymb}

\usepackage{lineno,hyperref}
\usepackage{cancel}
\usepackage{ulem}

\usepackage{graphicx}
\usepackage{dcolumn}
\usepackage{bm}
\usepackage{amsmath}

\begin{document}

\title{Bucket-brigade inspired power line network protocol for sensed quantity profile acquisition with smart sensors deployed as a queue in harsh environment}

\author{Edval J.\ P.\ Santos}
\affiliation{Laboratory for Devices and
Nanostructures -- Nanoscale Engineering Group,
Universidade Federal de Pernambuco, Recife-PE, Brazil\\
E-mail: edval@ee.ufpe.br or e.santos@expressmail.dk.
}

\begin{abstract}
Pressure and temperature profile are key data for safe production in oil and gas wells.   In this paper, a bucket-brigade inspired sensor network protocol is proposed which can be used to extract sensed data profile from the nanoscale up to kilometer long structures.  The PHY/MAC layers are discussed.  This protocol is best suited for low data rate exchanges in small fixed-size packets, named buckets, transmitted as time-domain bursts among high-precision smart sensors deployed as a queue.   There is only one coordinator, which is not directly accessible by most of the sensor nodes.  The coordinator is responsible for collecting the measurement profile and send it to a supervisory node.  There is no need for complex routing mechanism, as the network topology is determined during deployment.   There are many applications which require sensors to be deployed as a long queue and sensed data could be transmitted at low data rates.  Examples of such monitoring applications are: neural connected artificial skin, oil/gas/water pipeline integrity, power transmission line tower integrity, (rail)road/highway lighting and integrity, individualized monitoring in vineyard or re-foresting or plantation, underwater telecommunications cable integrity, oil/gas riser integrity, oil/gas well temperature and pressure profile, among others.   For robustness and reduced electromagnetic interference, wired network is preferred.  Besides in some harsh environment wireless is not feasible.  To reduce wiring, communications can be carried out in the same cable used to supply electrical power. 
\end{abstract}

\pacs{}

\keywords{measurement profile, smart sensor, network protocol, bucket-brigade, wired, power line communications.}

\maketitle

\section{Introduction}
\label{sec:Introduction}

\begin{figure} [t] 
\centering
 \includegraphics[width=0.51\linewidth]{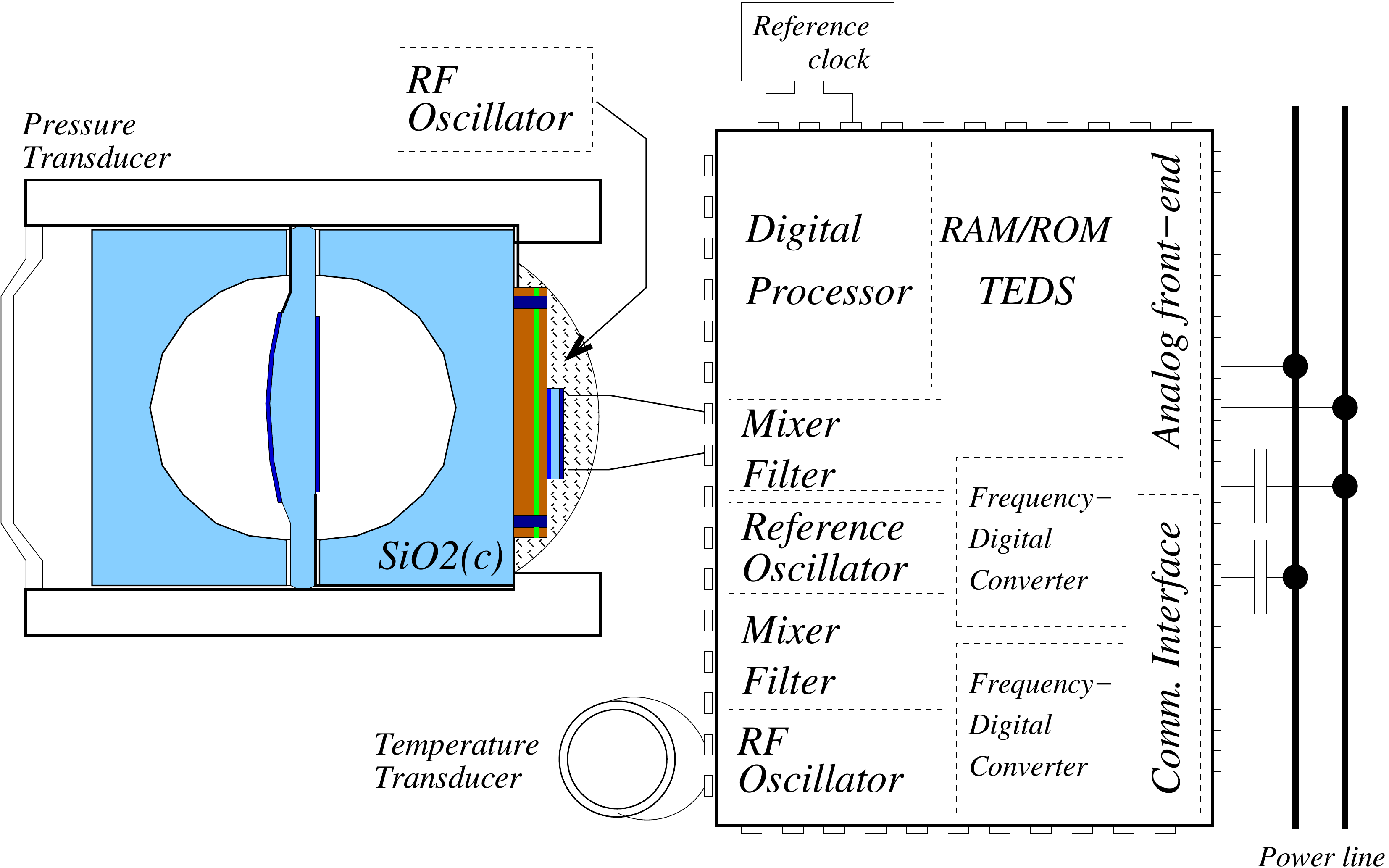} 
\caption{Smart sensor for pressure and temperature profile measurement in oil and gas well.}
\label{fig:smartsensor}
\end{figure}

Electronic sensors are highly flexible as they can be used to measure any physical or chemical quantity.  Besides, they can be designed to include data processing and data communications into a single chip, as shown in Figure~\ref{fig:smartsensor}.  Thus, network-capable smart sensors have become ubiquitous~\cite{IEEE1451,FFMST2009}.

Over the last 50 years, there has been a huge development in network technology for distributed measurement.  There have been many efforts to carry-out distributed measurement over different types of networks.  One approach is to implement as a protocol in the application layer of the OSI model~\cite{CA1988,BFOP1998,GM2001,PMC2003}.  This approach can achieve good performance for computer-based virtual instrumentation in local area network, but it is sensitive to electromagnetic interference, does not use the channel bandwidth efficiently, is not real-time and cabling costs is an issue.

For industrial applications, there are many wired network technologies, such as:  AS-Interface, CAN, Data highway 485, DeviceNet, EIA/TIA 232,  EIA 485,  Foundation fieldbus, HART,  Industrial real-time ethernet, Modbus,  Profinet/Profibus~\cite{FFMST2009,MWPR2004,YY2016, MPDO2010}.  Such network technologies are designed for local area networks and the cable technology used can be expensive.  To reduce cabling cost impact, there has been initiatives to develop wireless communications protocol for industrial environment~\cite{ASSC2002}.  One example is Security Equipment protocol in Routing in Ad-hoc Networks, SERAN~\cite{BFVGS2005,PPM2008}.  In some applications, such as: Vehicular Ad-hoc Networks (VANET), farm made up of several greenhouses, or geographically distributed devices, the communications link has to be wireless or hybrid wired/wireless~\cite{HVB2007,NFL2018,MB2011}.  

\begin{figure} [t] 
\centering
 \includegraphics[width=0.51\linewidth]{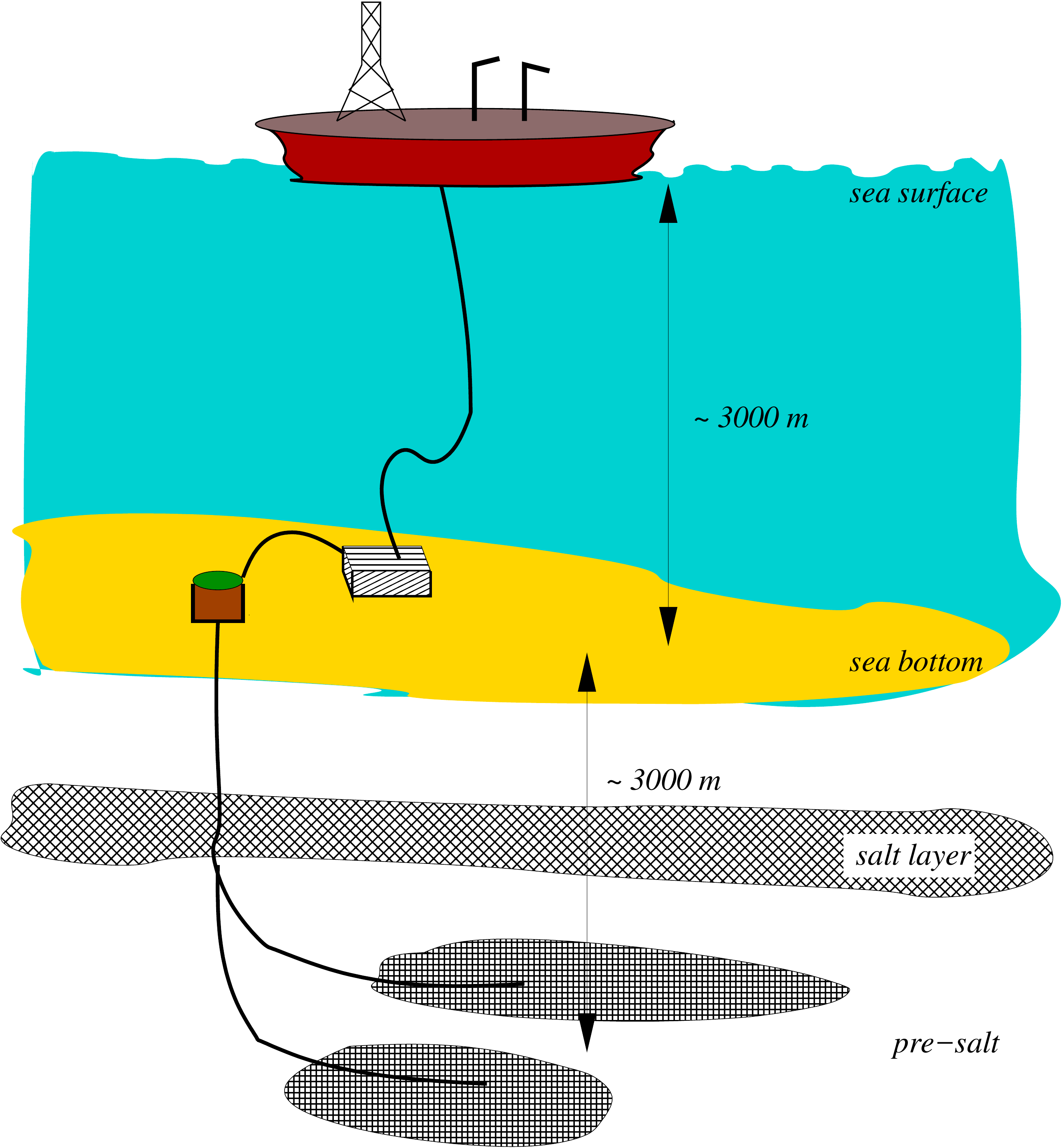} 
\caption{Floating production storage and offloading, FPSO, for oil extraction in the pre-salt.}
\label{fig:FPSO}
\end{figure}

The network protocols for sensed data acquisition are commonly designed for the local area network.  For the applications under consideration, a network protocol for sensed quantity profile acquisition is still missing.  Thus, the {\it bucket-brigade} sensor network protocol is proposed.   This protocol is best suited for low data rate exchanges in small fixed-size packets, named buckets, transmitted as bursts among high-precision smart sensors deployed in kilometer-long queue.     The {\it bucket-brigade} concept has been used previously in the development of algorithms for classifier systems in rule-based artificial intelligence and order picking in warehouse~\cite{credit,warehouse}.  The motivation to develop the {\it bucket-brigade} protocol started during the development of sensors for distributed measurement in oil production wells in the pre-salt layer~\cite{Santos2020,SS2019,SV2014,SV2008,SS2013b,SS2010,SS2013,LBMS2014,BR1020170153630}, as shown in Figure~\ref{fig:FPSO}.

Considering large structures submitted to harsh environment, there is a need for a network technology for smart sensors to carry out long haul measurement quantity profile acquisition for monitoring and surveillance.  Examples are:   neural connected artificial skin, oil/gas/water pipeline integrity, power transmission line tower integrity, (rail)road/highway lighting and integrity,  individualized monitoring in vineyard or re-foresting or plantation, underwater telecommunications cable integrity, oil/gas riser integrity,  oil/gas well temperature and pressure profile, etc.   The network can be homogeneous, in which all sensor nodes are identical, or heterogeneous.  In such applications is important to supply some synchronization service and the sensor nodes must be robust (safe and secure).  

Although for some of the sensor network examples in this paper the network could be wireless, in other application examples given, wireless technology cannot be deployed as it is too costly or signal transmission cannot be achieved.  Besides, typical wireless sensor network technologies are short range as compared to the network length discussed in this paper.  For robustness (safety and security) and to reduce electromagnetic interference, EMI, wired network is preferred.   However, for the long haul sensor network examples presented in this paper, cable deployment can be costly. Thus, it is preferable to use single cable to supply electric power and for data communications.  Power line communications, PLC, is a technology under development for over a century and is becoming widely used for home automation, home security,  computer network, telephone communications, remote home monitoring,  internet-of-things, gaming, audio-video streaming, solar energy inverters, electric car smart charging,  smart metering or remote meter reading, and smart cities. Technologies such as: HomePlug, ISO/IEC 14908, ITU T G.hn, IEEE 1901 have received wide acceptance in the home area network and for utility telemetry.  For long haul PLC, there is Two Way Automatic Communication System, TWACS.  But, it is ultra narrow band with very low data rate, up to 100/120 bps depending on mains power frequency.   For higher speed, PLC uses Quadrature Amplitude Modulation, QAM, combined with Orthogonal Frequency Domain Multiplexing, OFDM, schemes to send data over the electric cable at high data rates.  This  modulation technique  has been successfully used in both wired and wireless network protocols.  QAM is a combination of Phase Shift Keying, PSK, and Amplitude Shift Keying, ASK, techniques.   It is used in HomePlug, HD-PLC, IEEE 1901, ITU T G.hn, WLAN, 4G and 5G networks.  OFDM is robust against impulse or burst noise, narrow band interference and can avoid usage of frequencies with poor channel response or demanded by regulations.  But, it is sensitive to phase noise, frequency offset and timing errors~\cite{homeplug,LNLKY2000,OVLR2018,IEC14908,ITUG.hn,IEEE1901, CPMLTD2016,CKYK2010,HDW2004,PJSE2014}.

As oil production wells are becoming electrified, this offers an opportunity for deployment of smart sensor network as discussed in this paper.  Thus, to present further details of the proposed network protocol, a homogeneous oil well temperature and pressure  smart sensor network for profile measurement is used as application example.  The concept can be extended to the fishbone network topology, and can also be applied to nanosensors to mimic neural data transmission, however.  This paper is focused on the ISO/OSI layer 1 - PHY layer and the ISO/OSI layer 2 - MAC layer~\cite{ISO/OSI} and divided into six sections.  This introduction is the first.  Next,  the ISO/OSI layer 1 - PHY layer data processing blocks are described for the transmitter and the receiver.  In the third section, ISO/OSI layer 2 - MAC layer is presented.  In the fourth section,  the bucket-brigade protocol is presented.  In the fifth section, measurement of the oil well temperature and pressure profiles is discussed.   Finally, the conclusion.

\section{ISO/OSI layer 1: PHY layer}

The MAC Protocol Data Unit, MPDU, can contain measurement data with timestamp, beacon, management or acknowledgment.   The MPDU is the payload for the ISO/OSI layer 1,  PHY layer~\cite{ISO/OSI}.     It is transferred to the PHY layer for transmission over the communications link.  For {\it bucket-brigade} protocol, the selected channel encoding technique is QAM/OFDM, as this is a field proven technology used for PLC communications.  The PHY layer is responsible for encryption, error correction, mapping into OFDM symbols, generation of time-domain bursts~\cite{CPMLTD2016,CKYK2010,HDW2004,PJSE2014}.

In the PHY layer, the MPDU becomes, the PHY Service Data Unit, PSDU, which is prepared for transmission by adding preamble,  start of frame delimeter, SFD, to signal end of preamble, frame control with info such as frame length, resulting in the  PHY Protocol Data Unit, PPDU.  The PPDU preamble is a sequence of known symbols used for automatic gain control, synchronization, phase reference recovery.

\subsection{Bucket}

\begin{figure} [t] 
\centering
 \includegraphics[width=0.7\linewidth]{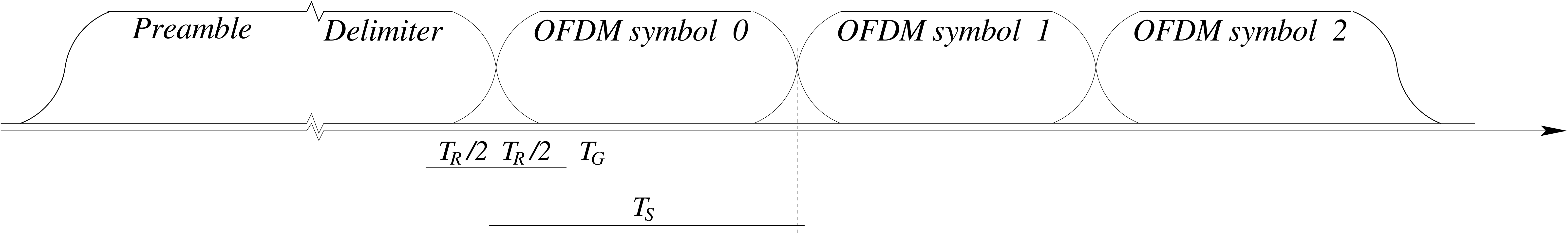} 
\caption{Bucket timing including preamble and three OFDM symbols with cyclic prefix. The prefix interval is a combination of the rolloff interval, $T_R$, and the guard interval, $T_G$.}
\label{fig:OFDMtiming}
\end{figure}

The PSDU converted into a time-domain OFDM signal burst to be injected into the channel is the bucket.   It is a fixed length time-domain frame.  In the oil well example, it is specified to contain 10 OFDM symbols as preamble and delimiter, plus 3 OFDM symbols carrying the same information as payload.   This bucket specification is not intended for audio/video streams.  Its timing is shown in Figure~\ref{fig:OFDMtiming}.   As the bucket size is fixed, the frame length field is not used.  For successful communication, clock frequency estimation may be required.  

\begin{figure} [t] 
\centering
 \includegraphics[width=0.51\linewidth]{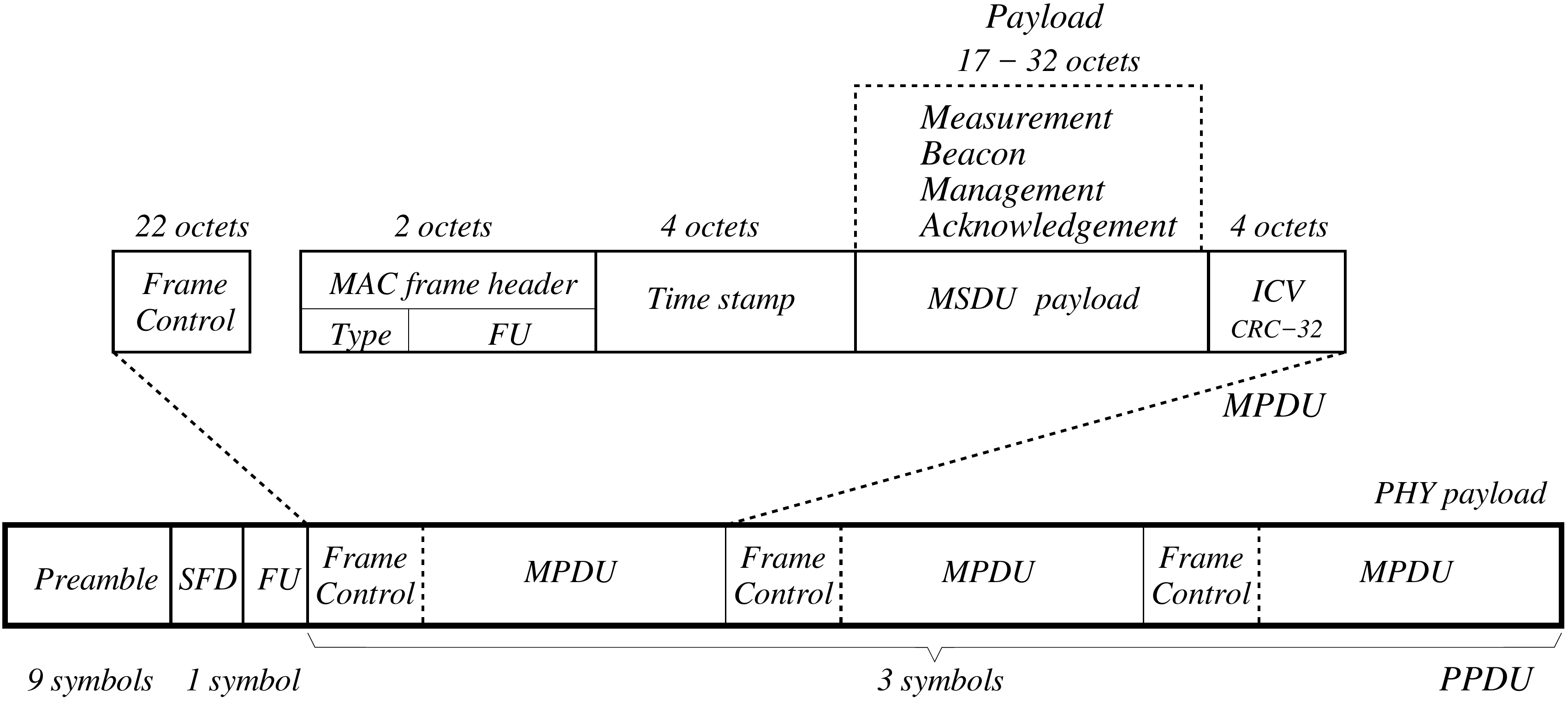} 
\caption{MAC protocol data unit, MPDU, and the PHY protocol data unit, PPDU.  SFD= Start of Frame Delimiter, FU= Future Use (e.g., frame length, sequence number, version).}
\label{fig:bucket}
\end{figure}

For fault tolerance, it can be specified to contain three replicas of the information payload (MPDU), as shown in Figure~\ref{fig:bucket}.  In case of three replicas, the coordinator can carry out majority vote.  Another possibility to improve fault tolerance is repeating the message in different subcarrier sets of a single OFDM symbol.

\subsection{PHY Service Data Unit, PSDU}

\begin{figure} [h] 
\centering
 \includegraphics[width=0.33\linewidth]{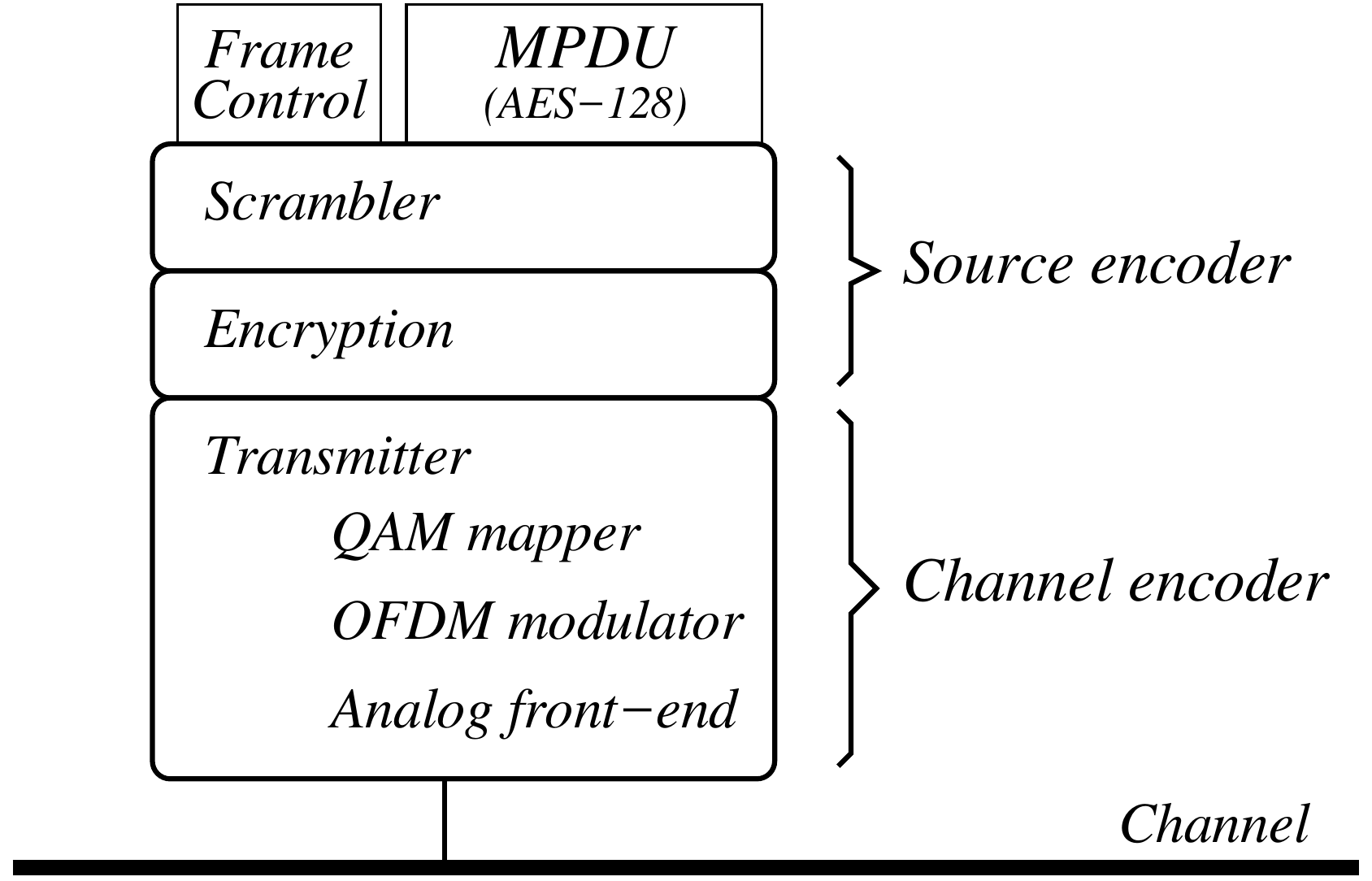} 
\caption{PHY blocks to create a bucket from the MAC Protocol Data Unit, MPDU.}
\label{fig:PHYblocks}
\end{figure}

Upon receiving the MAC Protocol Data Unit, MPDU, the digital sequence pass the scrambler. Next, encryption is performed.   Finally, the digital sequence, DS, is ready for QAM modulation, OFDM symbol construction, and time domain waveform generation, as illustrated in Figure~\ref{fig:PHYblocks}.  The receiver carries out the steps in reverse, as discussed in \ref{app:OFDM}.

\subsubsection{DS scrambler}
The data sequence is scrambled with a pseudo-random sequence to reduce the effect of impulse or burst error.  The polynomial used in HomePlug~\cite{homeplug} for DS scrambling is

\begin{equation}
  S(x)= 1 + x^3 + x^{10}.
\end{equation}

\subsubsection{Encryption}
To further protect data against burst or random errors, the digital sequence is transferred to the Forward Error Code, FEC, block for encryption.  To correct for random and burst errors, a combination of algebraic coding or block coding, which is good for burst errors, and convolutional code, which is good for random errors, is used~\cite{VO1979}.  Such combination is named concatenated code.  Between code blocks, an interleaver is introduced~\cite{homeplug}. 

The ratio of the original DS length, $k$,  of useful or non-redundant info to the sequence length with extra bits, $m$, due to FEC is defined as code rate, CR.  Thus, $m-k$ is redundant info used to correct errors. Typical code rates are as follows,

\begin{equation}
  \mbox{CR}= {k\over m}= {1\over 2}, {2\over 3}, {3\over 4}, {5\over 6}, {7\over 8}.
\end{equation}

HomePlug applies FEC 1/2 in Robust mode (ROBO).  This mode is used for transmission of beacon, data broadcast, multicast, management, and session setup.  For each 1 bit of information, there is an extra bit of code.  For other modes, HomePlug uses FEC 2/3.  Multiplying both numbers by 8, one gets $16/24$.  After puncturing,  FEC 2/3 code becomes $16/21$ or even $16/18$~\cite{homeplug}.  For the proposed bucket-brigade protocol, FEC 1/2 is preferred, as used in HomePlug ROBO mode.  Puncturing is not applied.

\section{ISO/OSI layer 2: MAC}

ISO/OSI layer 2 formats data frames into fixed length entities, performs channel access, and carries out error free transmission with automatic repeat request, ARQ~\cite{OVLR2018,CPMLTD2016,ISO/OSI}.  During configuration, the MAC frame or bucket can be set from 27 to 42 octets long.  For security, MPDU could be encrypted with AES-128.

{\small
\begin{itemize}
\item MAC header: 2 octets.
  \begin{itemize}
  \item Type (2 bits).
  \item Future use  (14 bits).
  \end{itemize}
  \item Timestamp: 4 octets (32 bits).
\item MAC payload: 17 -- 32 octets.
  \begin{itemize}
  \item Original source address: 4 octets (32 bits).
  \item Final destination address: 4 octets (32 bits).
  \item Current destination address: 4 octets (32 bits).
  \item Current source address: 4 octets (32 bits).
    \item Types of MSDU: 1 -- 16 octets (8 -- 108 bits).
    \begin{enumerate}
\item Measurement data.
\item Beacon.
  \begin{itemize}
  \item Central.
  \item Discover. 
  \end{itemize}
\item Management.
  \begin{itemize}
  \item Command/status type.
  \item Command/status payload.
  \end{itemize}
\item Acknowledgment.
  \begin{itemize}
  \item No payload (padding)
  \end{itemize}
  \end{enumerate}
\end{itemize}
\item Integrity check value (ICV): CRC-32 (4 octets).
\end{itemize}
}

Buckets types are: measurement data, beacon, management, and acknowledgment.  Possible commands and status info are:  acknowledgment, association start, association request, association end, automatic repeat query, coordinator alive and ready, data available, dissociation, global data request, global status request, hello neighbor, node data request,  node status request,  malfunction, node active, node failure detected, node address table request, neighbor inactive, orphan, perfect, priority bucket available, ready to retransmit, request sensorID/NodeID/NetID, set NodeID/NetID.

For error detection, cyclic redundancy check, CRC, can be applied and introduced in the Integrity Check Value, ICV, field.  The original DS is split into octet blocks.   Next, to calculate the CRC-$n$ code,  $n-1$ zeroes is appended to the DS.   The new sequence is $\mbox{mod}\,2$ (binary) divided by the CRC-$n$ generator polynomial, $G_{CRC}(x)$.  The remainder of the operation is the CRC code, which is appended to the message, replacing the zeroes appended before.

This binary division operation can be represented as a binary polynomial operation, $(P(x)/G_{CRC}(x))\mbox{mod}\,2$, in which, $P(x)$ is the original data sequence with $n-1$ zeroes appended.  One example is parity calculation, known as CRC-1.  One can verify that the CRC-$n$ generator polynomials, $G_{CRC}(x)$, are not divisible by $x$, but are divisible by $1+x$.  HomePlug uses CRC-32.  

{\small
\begin{equation}
  \begin{array}{ll}
  \mbox{CRC-1}\,\mbox{(Parity):} & 1+x\\
  \mbox{CRC-8}\,\mbox{(General):}& 1+x^2+x^4+x^6+x^7+x^8\\
  \mbox{CRC-8}\,\mbox{(ATM):}& 1+x+x^2+x^8\\
  \mbox{CRC-16}\,\mbox{(IEEE802.15.4):}& 1+x^5+x^{12}+x^{16}\\
  \mbox{CRC-32}\,\mbox{(IEEE802.3):}& 1+x+x^2+x^4+x^5+x^7+x^8+\\
  &x^{10}+x^{11}+x^{12}+x^{16}+x^{22}+\\
    &x^{23}+x^{26}+x^{32}
    \end{array}\nonumber
\end{equation}
}

\begin{figure} [t] 
\centering
 \includegraphics[width=0.51\linewidth]{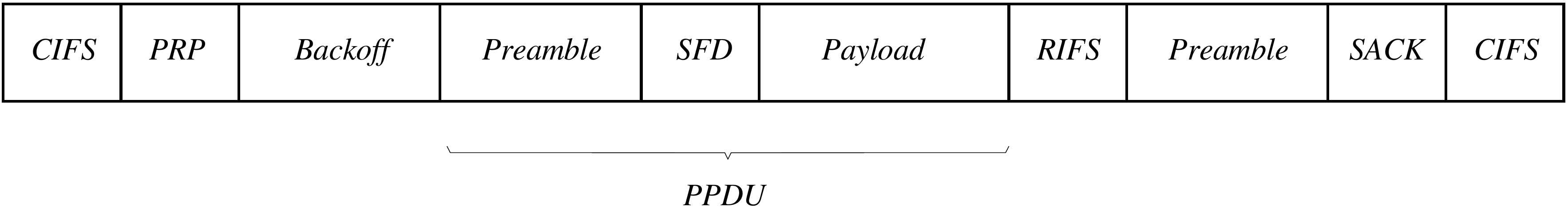} 
\caption{CSMA operation during collision.}
\label{fig:CSMA}
\end{figure}

Carrier Sensing Medium Access/Collision Avoidance, CSMA/CA, is the selected channel access method for the proposed {\it bucket-brigade} protocol.  Differently from HomePlug which uses a combination of CSMA/CA and Time Division Medium Access - TDMA synchronized to the mains frequency~\cite{homeplug,PJSE2014}.  In case of collision, nodes wait a backoff period.  First there is a Contention Interframe Spacing, CIFS, then Priority Resolution Period, PRP, and Backoff period.  After that the node tries to transmit the packet again.  Next comes the Response Interframe Space, RIFS, the Selective Acknowledgment, SACK, and Contention Interframe Spacing, CIFS, as in Figure~\ref{fig:CSMA}.  This results in overhead, lowering the datarate~\cite{homeplug,HPLL2016}.

The measurement cycle is a beacon-like period not necessarily synchronized to the power line signal, and status/command transmission is CSMA/CA period.  During measurement cycle data exchanges follows a sequence as in a bucket-brigade which allows for almost contention free data transmission and real-time constraint.   In most applications, bucket is a small fixed-length packet transmitted over a homogeneous network, in which there is no need for fragmentation.

\section{Bucket-brigade protocol}

Bucket brigade transfer method has been observed in ants ({\it Messor Barbarus}), equipment assembly, food preparation, order picking in warehouses, rule-based classifier systems in artificial intelligence, neural nets, and fire brigade~\cite{credit,warehouse}.

\begin{figure} [t] 
\centering
 \includegraphics[width=0.51\linewidth]{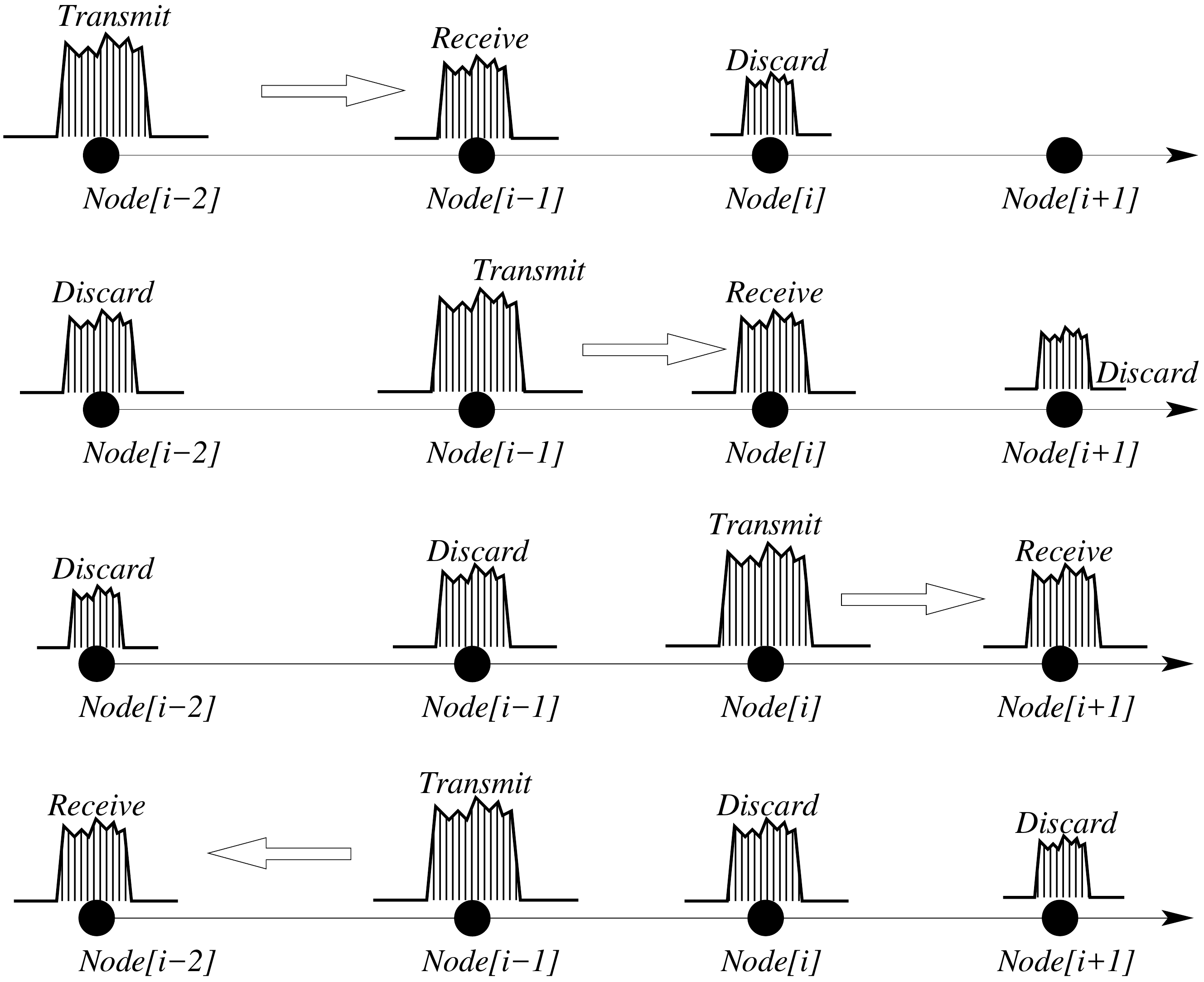} 
\caption{Message hopping or relay as bucket passing in a bucket brigade.  If a bucket is not intended to the node, it is discarded.}
\label{fig:message}
\end{figure}

In this paper, the concept is applied to create a network protocol for sensors deployed as a long queue.  For the application of this protocol, each node is also a repeater.  The idea is that each node sends its collected data in a packet, named bucket, to its next upstream neighbor, until it reaches the coordinator.   The bucket is a time-domain signal burst.  The bucket passing process is illustrated in Figure~\ref{fig:message}.   It is a message hopping or relay mechanism.

 The {\it bucket-brigade} network protocol is somewhat like token passing schemes or message relay. Examples of token passing network protocols are Token Ring,  ARCNET (Attached Resource Computer Network), and FDDI (Fiber Distributed Data Interface).  Token passing is a channel access method to avoid collision, in which the node has to get a token to send a message.  It is based on a ring configuration that can be formed physically or logically.  One node is the ring monitor, but in case of monitor failure, all nodes are equivalent and can assume the ring monitor role~\cite{IEEE802.5}.  ITU T G.hn uses an implicit token mechanism~\cite{ITUG.hn}.    Differently from token passing networks, the proposed {\it bucket-brigade} network does not form a ring, neither physically nor logically.  It is a centralized topology.  There is only one coordinator and sensor nodes can not assume the coordinator role.   The {\it bucket-brigade} protocol uses a form of message relay, which is used in the internet and in peer-to-peer or mesh networks, such as ZigBee~\cite{SF2008}.   Unlike mesh networks, the proposed network does not need to carry out routing discovery.  All nodes are capable of performing message relay and the topology is known.

\subsection{Network formation and operation}

The basic characteristics of the bucket-brigade network are as follows:

\begin{itemize}
\item There is only one active coordinator.
\item Sensor nodes are deployed as a queue.
\item Sensor nodes cannot become coordinator.
\item All sensor nodes are always listening to the network.
\item All sensor nodes are capable of message relay.
\end{itemize}

\subsubsection{Coordinator}

The coordinator is the master node which acts as network server and router.  All network traffic between the external world and the sensor network is carried out through the coordinator.  Hence, the coordinator must be resistant to cyber attacks.   Besides, it may be necessary to include slave coordinators to avoid the single point of failure problem.  Periodically, the coordinator sends a message to the slave coordinators stating it is alive.  In case of coordinator failure, a slave coordinator previously on-line has to assume its role.  Thus, the slave coordinator is always active and logging all network traffic to resume where the master coordinator has stopped, if required.   Security can be implemented at the coordinator node only or at each node, if the required electronic complexity can be realized.  To reduce sensor node circuit complexity, real time is kept by the coordinator node only.  The coordinator provides synchronization service, as it can issue a global write with a timestamp.  It also acts as a bridge to provide co-existence with other network technologies.

\subsubsection{Sensor node discovery and association}

\begin{figure} [t] 
\centering
 \includegraphics[width=0.51\linewidth]{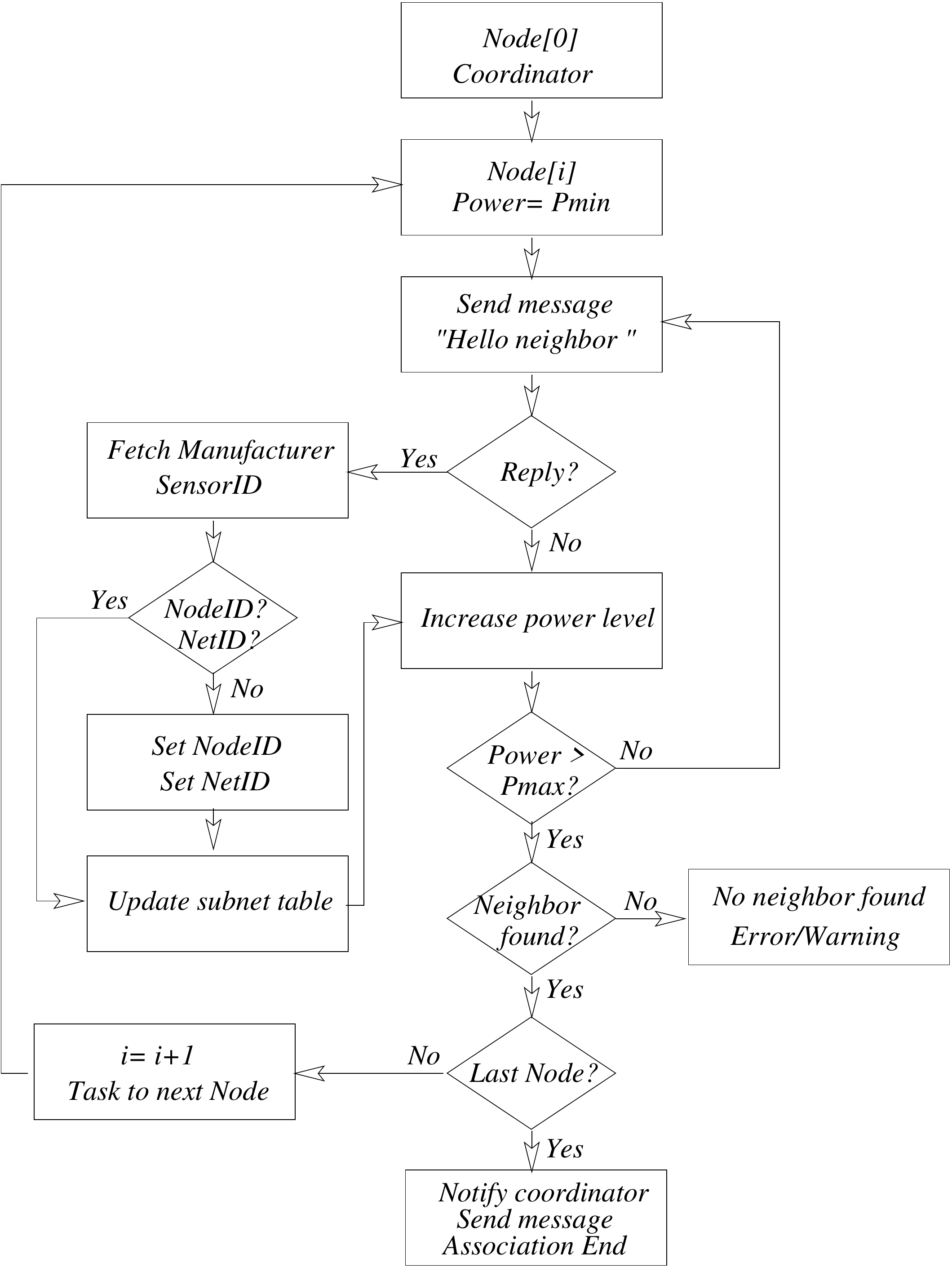} 
\caption{Flow diagram to illustrate the automatic sensor node discovery and association process in sensor network deployed as a queue.}
\label{fig:Association}
\end{figure}

To achieve Plug\&Play address assignment and configuration, each sensor node has a unique sensor ID attached to it by the manufacturer, and stored in its Transducer Electronic Datasheet, TEDS~\cite{IEEE1451}.   Initially, no sensor node has assigned node address (NodeID) or network address (NetID).  During network deployment, NodeID and NetID can be assigned either manually or automatically.  Automatic assignment could be achieved with sensor nodes capable of varying their output power level.  The procedure for automatic assignment is outlined in Figure~\ref{fig:Association}.

Starting from the lowest power level, the coordinator sends a bucket with a ``Hello neighbor'' message for node discovery.  It will increase the power level until it gets the first reply.  As the nodes are deployed as a queue, the first reply should come from the closest node.  This node gets the sensor ID/NodeID/NetID request.  If address is not set, it receives the set NodeID/NetID command.  This procedure will continue until the maximum power level is reached and no new sensor nodes are found.  Next, the coordinator authorizes the first configured node to repeat the procedure to find new nodes downstream.  If a node has found at least one neighbor, but has reached its maximum output power, it checks the neighbor table and signals the next neighbor downstream to start the association process. If a node has already replied a ``Hello neighbor'' request from an upstream node, it will keep silent/idle for further requests. This process continues until there is a node that is unable to get any reply from downstream neighbors.  This is assumed to be the last node.  This last node will issue association-end bucket to the coordinator.  After the association-end bucket, all newly configured nodes could send a ``node active'' message to the coordinator through its neighbors to confirm it is ready to collect data.  Sort of DHCP service.  All nodes stays in idle mode, with the receiver block listening to the channel.

There is no need to keep complex routing table.  After association all nodes contains a table of reachable nodes and minimum power level for each reachable node, but will only carry out communications with the closest neighbor.   Other reachable neighbors are only contacted in case of next neighbor failure to respond.   Thus, one criteria to improve fault tolerance is to establish that each node should be capable to contact at least two neighbor nodes upstream and at least two neighbor nodes downstream.  Should the failure be temporary, the node becomes orphan.  In case of node failure detection, the coordinator may issue a warning to the supervisory station.  The orphan node can send an association request bucket to the coordinator, which in return can start a new association procedure or check the older table for the requesting orphan node original neighbors and request them to carry out a table update to include the orphan.  In any case, node failure or orphan status requires the coordinator update its list of active nodes.  This can be carried out during network self-test phase.

\begin{figure} [t] 
\centering
 \includegraphics[width=0.33\linewidth]{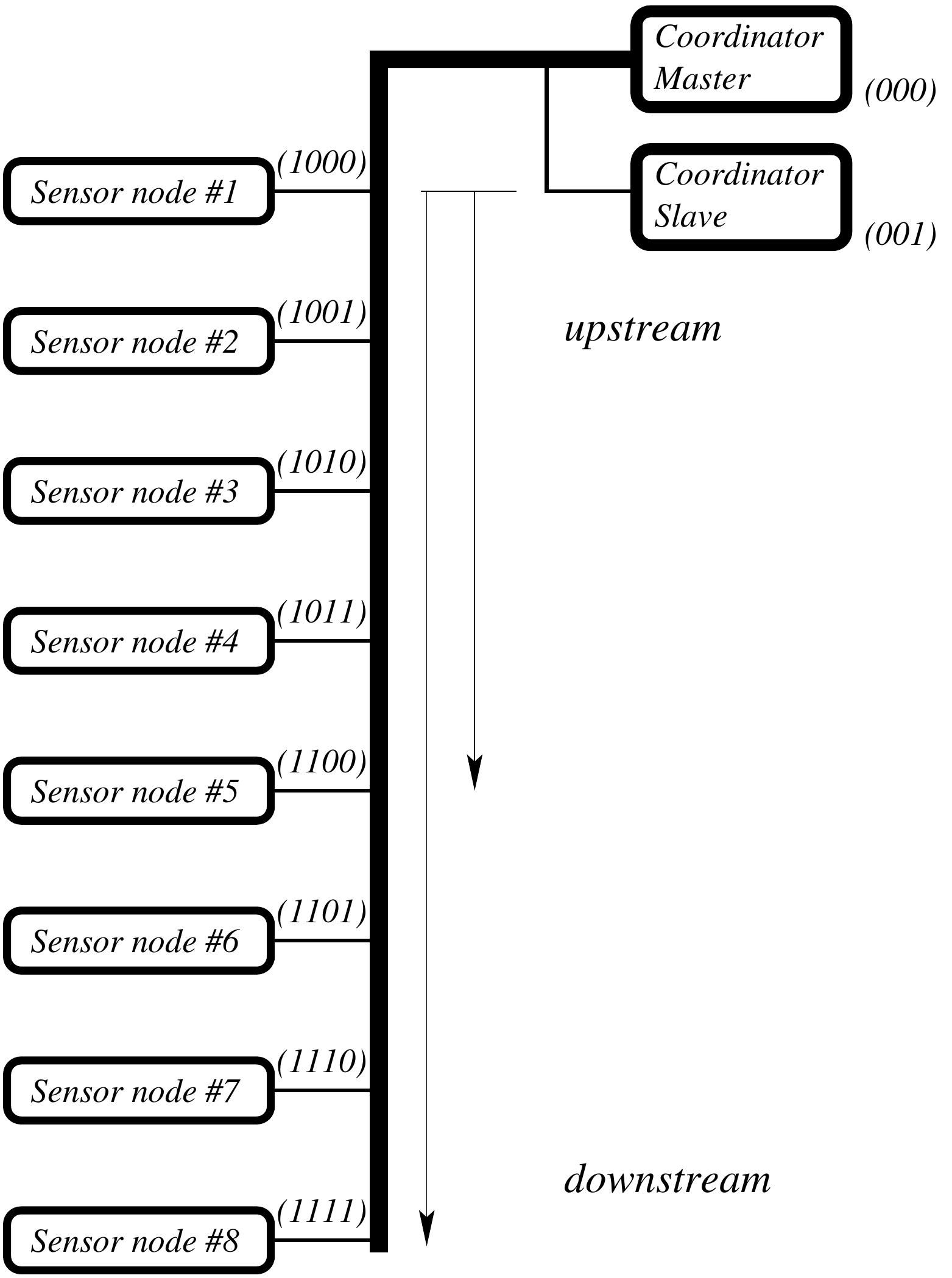} 
\caption{Smart sensor network with 8 nodes after association process is completed and NodeID and NetID are assigned.}
\label{fig:smartnetwork}
\end{figure}

The physical network can be arranged as a single line or as multiple lines, as a fishbone.  In case of bifurcation, GPS-like signal could be used to help with network assignment.  In case GPS-like signal cannot be used, the coordinator can  interpret the routing tables by checking NodeID/NetID and sensor ID associated with it, and send a message reassigning NodeID/NetID for a specific group of nodes.  It may be necessary to repeat the association procedure for this specific group of nodes. Tasks could be implemented in an operating system, such as TinyOS~\cite{tinyos}.   An example of a deployed and configured smart sensor network with 8 nodes is shown in Figure~\ref{fig:smartnetwork}.

\subsubsection{Priority}

Messages from or to the coordinator have the highest priority.  Nodes within a subnetwork can interfere with each other.  Thus, some sort of coordination is required to minimize such interference.  If the channel is idle and no measurement cycle has been initiated, the node can transmit its bucket immediately to assert priority resolution.  If channel is not idle, the node waits for end of transmission.  It can be established that a measurement cycle has to finish before warning or other status messages sent by sensor nodes are transmitted.  During measurement cycle,  as in the bucket-brigade, requests from upstream node has priority over requests from downstream nodes, as measurement data transmission is initiated by the upstream node.   Event driven bucket should wait the end of the measurement cycle.  In event driven period, node from downstream node has priority over upstream node.  Should the queue present multiple lines, different branches can get priority assignment based on the NetID.  Priority can also be assigned based on sensor position in the network.  Priority could increase towards the most downstream node.

\subsection{Data acquisition and transmission}

In normal operating condition, all sensors nodes have assigned NodeID and NetID, know their next available neighbors, and are idle.  Data measurement cycle is always started by the coordinator as a periodic known traffic.  Coordinator broadcast a data measurement request bucket with timestamp.  Next, it starts a timer considering the latency for the lowermost node to reply.  A sensor node first transmits its data upstream, then requests data from its downstream neighbor.  The downstream node awaits listening for the upstream node data request addressed to it.   Thus avoiding contention during a measurement cycle.  Distant sensor nodes buckets are like interference signals.  This is like a beacon period for all nodes to reply as a bucket-brigade.   Thus, transmission schedule is determined by sensor position in the network.  All sensors should reply with measurement data with the timestamp transmitted by the coordinator, which keeps a record of all nodes which replied to the data request and which nodes are pending.  During normal measurement operation, the following procedure could be executed to avoid collision and to achieve a deterministic latency:

\begin{itemize}
\item All sensor nodes are idle and listen to the channel for carrier detection.
\item Coordinator node sends a data measurement request bucket with a timestamp to start a measurement cycle.  Next, it sets off a timer to wait for all nodes to reply.
\item Sensor node detects channel activity, communications module wakes up and checks for valid preamble, indicating message presence.
\item Decode source address and bucket type.  Discard if not for the node and return to idle.
  \begin{itemize}
  \item if data measurement request bucket from coordinator, sensor node passes the data measurement request bucket to its closest neighbor downstream, which repeats the process.
  \item if data measurement bucket is from downstream neighbor, repack to send to the next neighbor upstream, after its own data is transmitted.
  \item if event driven bucket is from downstream neighbor, repack to send to the next neighbor upstream after the measurement cycle ends.
  \item None of this, returns to idle mode.
  \end{itemize}
\item if data request from coordinator, acquire data and prepare bucket for transmission.  Thus, measurement is carried out in parallel.  
\item Node closest to the coordinator transmit first and fetches next bucket from its closest downstream node available.
\item Each node checks it has sent the message successfully and got an acknowledgment, ACK, reply from the upstream node, before listening to the node downstream for messages to send.
\item Buckets are passed up until the bucket from the last node at the bottom reaches the coordinator.
\item All sensor nodes go to idle mode again.
\end{itemize}

\begin{figure} [h] 
\centering
 \includegraphics[width=0.51\linewidth]{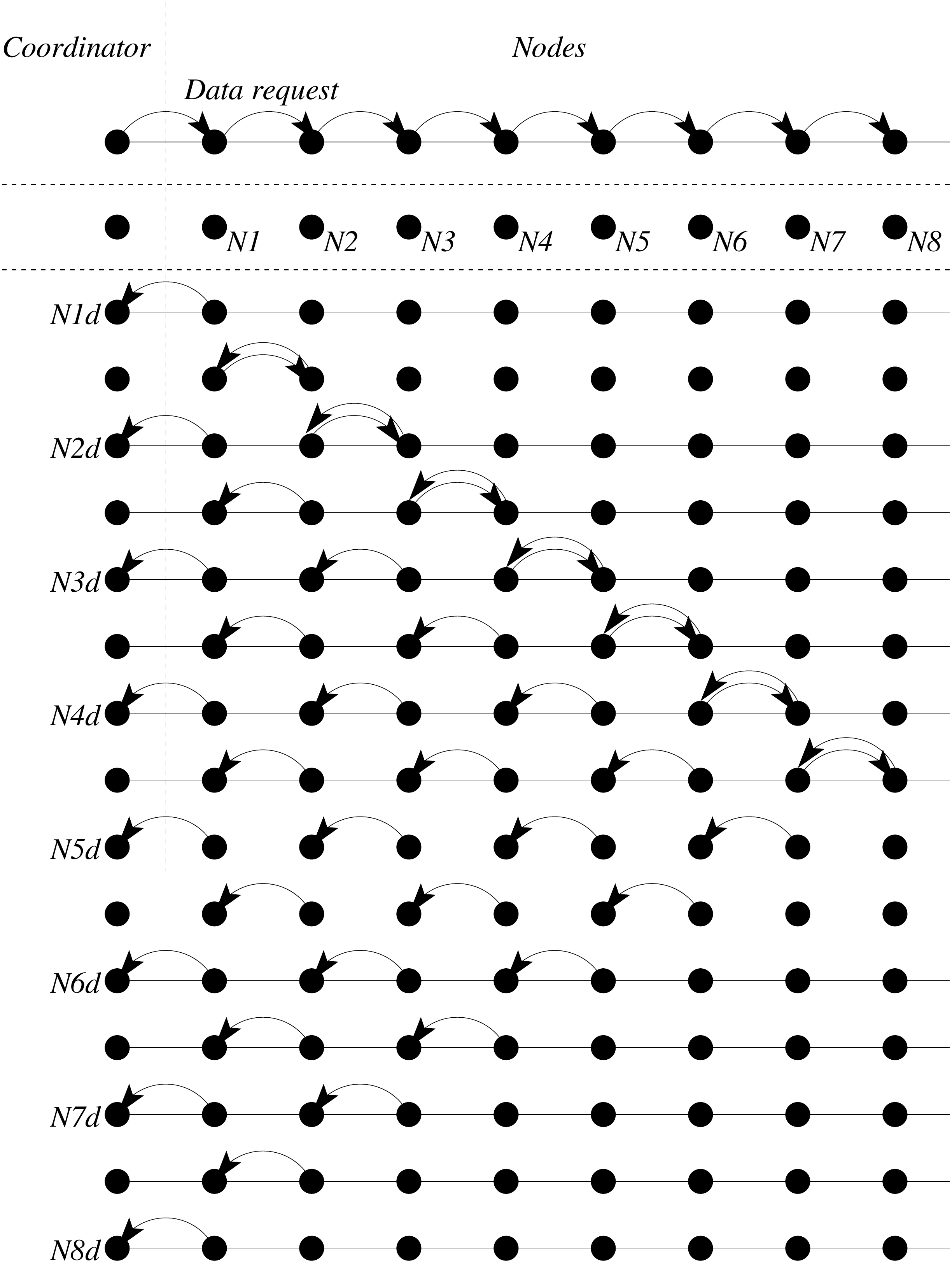} 
\caption{Data transfer as a bucket-brigade in a smart sensor network with 8 nodes.}
\label{fig:datatransfer}
\end{figure}

As bucket with data from last node downstream arrives at the coordinator, the measurement cycle ends.    The process is illustrated in the first row in Figure~\ref{fig:datatransfer} for a smart sensor network with 8 nodes.  Differently, from token ring network, the bucket-brigade smart sensor does not need to wait for a bucket to send a message.  It is capable of producing priority messages as warning messages, such as self-test failure or abrupt change or threshold crossing of a measured quantity.  During measurement cycle, status messages are halted.  In case of a heterogeneous network, direct coordinator-node communications can be used to send a long data sequence, such as audio or image, as a sequence of buckets.

\subsubsection{Latency}

To estimate an upper bound for the measurement process latency, assume the coordinator has started a measurement cycle.  Nodes are listening to the network and received the data measurement request bucket originated from the coordinator node.   As a given node receives the command to carry out a measurement, it re-assembles it and sends to the next downstream neighbor before starting its own data acquisition process.  Thus, the time the command takes to reach the last node is the number of nodes, $N$, multiplied by the time to transfer data between nodes, $T_{n-n}$, combined with the internal latency to process the command, $T_{IL}$,

\begin{equation}
  N(T_{n-n} + T_{IL})
  \label{eq:commanddown}
\end{equation}

Next, measurements are carried out in parallel during interval $T_{acq}$ and the bucket is built to be sent upstream.   For high resolution part-per-million measurement with a frequency to digital converter, acquisition time is about 1 second~\cite{SS2010}.  Node closest to the coordinator will transmit first.  After finishing its transmission, this node request data from the second closest.  Thus, each node transmits its measurement first to the upstream neighbor then listen to downstream neighbor node.   The coordinator records the time in which the packet from the last node arrives.   The time for the last node data bucket reaches the coordinator is:

\begin{equation}
 (2N-1)(T_{n-n} + T_{IL}) + T_{acq}
  \label{eq:measuretime}
\end{equation}
in which,  $N$ is the number of deployed nodes, $T_{n-n}$ is the data rate time between nodes, and $T_{acq}$ time to carry out a measurement.

For this passing bucket approach, combining Equations~\ref{eq:commanddown} and \ref{eq:measuretime}, latency is estimated as follows,

\begin{equation}
  \mbox{Latency}\approx (3N-1)(T_{n-n} + T_{IL}) +  T_{acq}
  \label{eq:latency}
\end{equation}

\subsection{Node failure}

During the measurement cycle, in case of lack of communication with a neighbor, a sensor node will try to contact the second closest neighbor listed in its active neighbor table.  This is a self-healing process.  In case of missing or erroneous message from a sensor node, the coordinator could select a specific sensor node to re-send the erroneous data.  Thus, node failure can be detected by the coordinator.  If a node fails to send its bucket within a given time period, the coordinator can carry out maintenance tasks, if needed. 

After completing the measurement cycle, the coordinator can wait a period for pending event driven buckets to arrive.  A sensor node which presents self-test failure or detects an inactive neighbor can inform the coordinator with a warning as event driven bucket.

\section{Application to the oil and gas well}

To present some of the numerology, pressure and temperature profile in oil and gas production well is considered~\cite{Santos2020}.   Temperature and pressure transducers can be mounted as part of oscillator circuits and measured data can be directly converted from frequency to digital~\cite{SS2010}.  Differently from typical industrial applications, sensor robustness and reliability is even more critical and has to be combined with sensor compactness for distributed measurement.  The network must be deployed to last over 10 years, as in a space application.

For the oil and gas well example, it is assumed there are 1000 sensor nodes with 10~m separation in a queue 10~km long.  It is also assumed a node should be capable of 10~dBm maximum output power with 15~dB signal-to-noise ratio at the transmitter output, and the receiver sensitivity is -50~dBm.  Based on Equation~\ref{eq:errorprob} and Figure~\ref{fig:SER}, 16-QAM is selected.  For more robust transmission 4-QAM could be used.

During the well completion phase, to carry out association, each node starts  -5~dBm output power and increase at 1~dB step until it gets a reply.  Thus, 4-bit resolution to control the output power should be enough,

\begin{equation}
  {15~dB\over 2^4}\approx 0.94~dB
\end{equation}

Measured power loss along the electric and communications cable is about 0.3\,dB/m.    As nodes are deployed with 10~m separation, and assuming error can be corrected down to 0~dB signal-to-noise ratio, hence, more than two neighbors to each side are reachable by every deployed node as signal can travel up to 50~m and still be recovered.  After well completion, each node has built its table of reachable neighbors with the power level required for reliable communication. 

In this application, there is no GPS-like signal to confirm sensor position.  The coordinator could request a temperature measurement from all nodes under no flow, and compare to the expected geothermal gradient to estimate node position.  Sensor address may be assigned indicating sensor priority.  For example, sensor closest to the well bottom is issued the highest priority in event driven communications.

Both, temperature and pressure, are physical quantities which can be represented in fixed integer notation.  For a maximum pressure of 10$^8$ Pa (1000\,atm), 27-bit integer ($8/\log2\approx 27$) can be used to represent the measured quantity, and for a maximum temperature is 10$^3$ K, a 10-bit integer  ($3/\log2\approx 10$) can be used.  Thus, 5 octets is enough to carry measured data.  As discussed in Section~III, the bucket can be set to use this number of octets during well completion.

Considering coaxial cable already used in oil and gas well applications, available bandwidth is over 25~MHz.  Smart sensor node communications can be carried out in the frequency range from 6.25~MHz up to 18.75~MHz, with subcarrier bandwidth of 48.828~kHz,

\begin{equation}
  {25.0\times 10^6\over 48.828\times 10^3}= 512
\end{equation}

Thus, for the frequency range from 6.25~MHz up to 18.75~MHz, power line communications is carried out with subcarrier number ranging from 128 to 384, yielding 256 for the total number of OFDM subcarriers. 

The transmitted symbol is extended with the guard interval or cyclic prefix.  For this example,  25\% cyclic prefix is used,

\begin{equation}
   T'= {1 + 1/4\over 48.828\,kHz}= 25.6\, \mu s
\end{equation}

The subcarrier bandwidth is $2/T'$.  Thus, according to Equation~\ref{eq:phyrate}, the expected maximum raw PHY data rate is

{\small
\begin{equation}
\mbox{PHY rate}= 40 \,Mbps
\end{equation}}
in which, $n= 4$ is from the $2^n$ QAM modulation, CR= 1/2 is the code rate.

It is assumed 16-QAM is the only available modulation, payload is split into 4-bit blocks prior to IFFT processing with the 256-point FFT block.  Should this not be possible 4-QAM could be used.

The bucket duration is $13 \times T'= 0.33\, ms$.  Thus, the expected maximum raw bucket rate, MRBR, is

\begin{equation}
   \mbox{MRBR}= {1\over 13\times T'}= 3004\, \mbox{Bucket}/s
\end{equation}

Sampling rate at the receiver is one sample per symbol times the amount of subcarriers.

\begin{equation}
  f_s= {N\over T}= N\times \Delta f= 256\times 48828= 12.5\, \mbox{MSPS}
\end{equation}

MPDU is set to 31 octets, FC is 22 octets, thus a symbol is 424 bits plus cyclic prefix.  For CR= 1/2, this is converted into 848 bits.  For 16-QAM, this requires 212 carriers from the 256 carriers available.  Thus, 44 carriers can be used to transport known symbols, named pilot symbols or pilot waves for channel estimation and equalization.  In this set, there are excluded subcarriers due to poor channel quality for the particular subcarrier or to avoid interference with other services.  A summary of the proposed sensor network specification is presented in Table~\ref{tab:netspecs}.

\begin{table}[t]
\caption{Specification example for the proposed bucket-brigade network to be deployed in the oil and gas production well.}
\label{tab:netspecs}
\centering
{\small
\begin{tabular}{|l|c|}
\hline
{\bf Property} & {\bf Value}\\
\hline\hline
Cable             & Coaxial cable \\
\hline
Channel bandwidth & 25 MHz\\
\hline
Subcarrier bandwidth & 48.828 kHz\\
\hline
Total number of subcarriers & 512 \\
\hline
Total OFDM subcarriers & 256 \\
\hline
Channel allocation    & k= 128 -- 384 \\
                      & (6.25 MHz -- 18.75 MHz)\\
\hline
Modulation & 16-QAM  \\
\hline
FFT        & 256 \\
\hline
Guard interval (cyclic prefix) & 25\%\\
\hline
Data rate (bucket/s) & 3,000  \\
\hline
Dataset profile  & $< 1\,$s\\
transmission time & \\
\hline
Encryption & AES 128-bit (optional)\\
\hline 
Forward error code & Turbo concatenated code\\
\hline
Code rate & 1/2 \\
\hline
Transmission power & -50 dBm -- 10 dBm\\
\hline
Output power step (4 bits) & 1 dB \\
\hline
Receiver sensitivity & -50 dBm\\
\hline
Power consumption &  20 mW\\
\hline
Operating temp. & -40\,$^o$C --- 300\,$^o$C\\
\hline
Pressure & $\le$1000~MPa\\
\hline
Packaging & Stainless steel\\
\hline
Voltage & 5~V\\
\hline
Current (sleep mode) & $<$10 $\mu$A \\
Current (idle mode) & $<$1~mA\\
Current (transmission) & 20~mA (0\,dBm)\\
\hline
\end{tabular}
}
\end{table}

According to Equation~\ref{eq:latency}, one can estimate the latency for 1000 Nodes.  Communication time between nodes is roughly the duration of the Bucket, $13 \times T'= 0.33\,ms$.  Assuming node internal processing time is one-tenth of the data transfer time.

{\small
\begin{equation}
  \mbox{Latency}=(3\times 1000 - 1)(0.33 + 0.033)\times10^{-3} +  1= 2.088\,s
\end{equation}
}

This latency value can be reduced by applying higher complexity QAM/OFDM communications.  In any case, in a few seconds the supervisory node can get valuable information about a very large structure.  In this example, 10~km long.

\section{Conclusion}

A protocol for long haul smart sensor network deployed as a queue is proposed.  It uses fixed-size message, named bucket, which is transmitted by message relay, as the coordinator is not accessible for most nodes.  The message transfer technique is inspired in a bucket-brigade.  Bucket transfers are carried out with the minimum power level possible to reduce contention.  Distant nodes are perceived as impulse noise.  As the bucket carries a fixed amount of OFDM symbols, it is simpler to achieve known latency condition, which can yield deterministic data transfer from the network.  The proposed protocol can be extended for larger data packets at higher data rates by applying higher complexity QAM/OFDM schemes.  The coordinator act as a bridge to allow for co-existence with other types of networks and protocols.

The characteristics of the proposed network are: queue topology, wired, compact smart nodes, length over 1~km, fault tolerant, redundancy, fail-safe operation, robust, low datarate, known latency for real time data acquisition and transmission, cyclic, self-diagnose.  This proposal can be easily extended for fishbone type network topology and for higher effective data rates and higher complexity messages, such as: audio and video stream, collected from specific nodes along the queue.  Protocol parameter values can be adjusted for optimization to a specific application.  This proposal is not a complete specification.  Further technical details are required.


\section*{Acknowledgments}

The author thank Eng. Maur\'{\i}cio Galassi for many fruitful discussions in the initial part of this investigation.

\appendix

\section{OFDM}
\label{app:OFDM}

\subsection{Transmitter}

OFDM allows for high-speed data transfer in a slow noisy channel.  The key innovation behind OFDM is that data transmission is carried out in parallel by using subcarriers, which are all orthogonal to each other, meaning that the integral of the product of any pair of subcarriers over the symbol period yields zero.  Thus, each subcarrier can be thought as a Fourier component, whose amplitude is modulated using QAM.  The simplest form of QAM is Binary Phase Shift Keying, BPSK, in which two 180$^o$ separated phases are used to generate the time domain signal, $s(t)$.  This is an one-bit modulation, also known as 2-PSK. 

\begin{eqnarray}
  s(t)&=& I(t) \cos(\omega_c t) + Q(t)\cos(\omega_c t + \pi/2)\\
      &=& I(t) \cos(\omega_c t) - Q(t)\sin(\omega_c t) \label{eqno:04}
\end{eqnarray}
in which, $I(t)$ is the in-phase signal, $Q(t)$ is the quadrature signal, and $\omega_c$ is the carrier frequency.

\begin{figure} [t] 
\centering
 \includegraphics[width=0.51\linewidth]{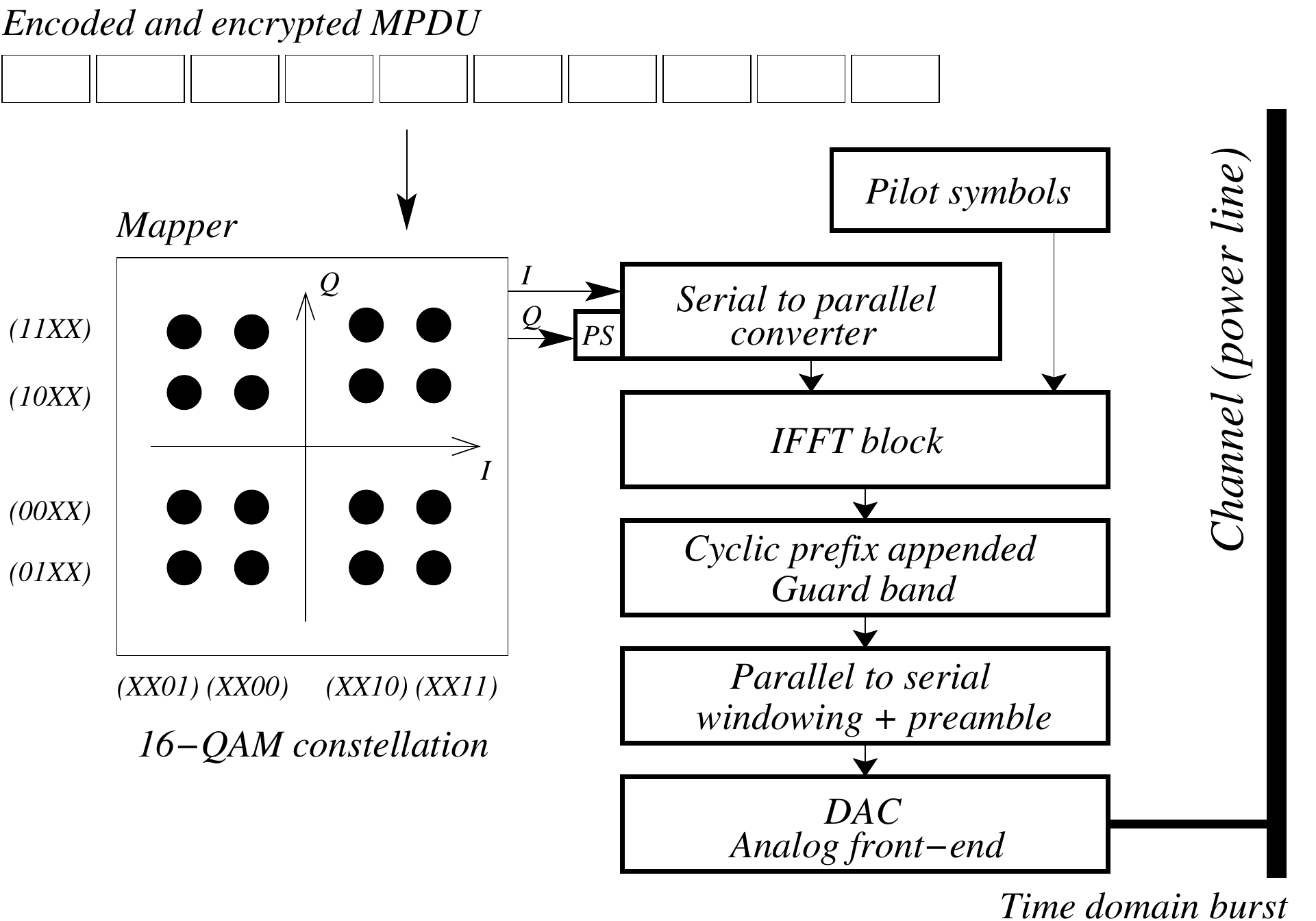} 
\caption{Block diagram of the QAM/OFDM transmitter.}
\label{fig:OFDMtransmitter}
\end{figure}

 In practical implementations, signals are synchronized and differential mode is applied.   A higher data rate scheme is the quadrature phase shift keying, QPSK, which is a two-bit modulation scheme, also known as 4-PSK or 4-QAM.  For larger number of bits, QAM is referred to as 2$^n$-QAM or M-QAM. Up to 8-QAM it is PSK only modulation.  At higher orders, it is a combination of PSK and ASK.  The set of QAM symbols is named constellation or alphabet.  The constellation can be indexed as ${\cal M}= [(2k - 1 - \sqrt{2^n}) + j(2l - 1 - \sqrt{2^n})]$, in which $(k,l)= 1,2,...,\sqrt{2^n}$.  Thus, typical values of M are: 16, 64, 256, 1024, 4096.  Certainly, more bits implies reduced symbol distance and increased susceptibility to error.  The QAM modulation can be changed depending on channel quality.  Thus, good link conditions allows for higher order modulation which translates into higher effective data rate.  The simplified transmitter block diagram is shown in Figure~\ref{fig:OFDMtransmitter}.  

HomePlug 1.0 uses coherent DQPSK, HomePlug 1.0 turbo adds 256-QAM, HomePlug AV adds 1024-QAM. HomePlug AV2 adds 4096-QAM.   ITU T G.hn uses FFT-OFDM with up to 4096-QAM modulation.  In HomePlug 1.0, 1024-point FFT is used and for HomePlug AV2, it is increased to 8192-point FFT~\cite{homeplug,LNLKY2000}.  

The MPDU is split into N blocks with $n$ bits.  Each $n$-bit block,  is encoded in 2$^n$-QAM becoming a complex number, $d_k= I_k + jQ_k$ ($k=0...N-1$). The in-phase, $I$,  and the quadrature, $Q$, can be thought as axis of a complex plane.  Thus, a symbol is a point in the $I\times Q$ plane.  Each obtained QAM symbol, $d_k$, modulates an orthogonal subcarrier, as in Equation~\ref{eqno:04}, and carries $n$ bits of the original message.  

The spectrum, $S_l(f)$, is the sinc function with bandwidth $2/T$, $l= 0...L-1$ is symbol index. 

\begin{equation}
  S_l(f)=  \sum_{k=0}^{N-1} d_{k,l} \,{\sin\left(\pi\left(f-{k\over T}\right)T\right)\over \pi\left(f-{k\over T}\right)T}\label{eq:05}
\end{equation}
in which,  $d_{k,l}$ is a QAM modulated symbol for the $k$-th subcarrier in the $l$-th symbol, $k=0...N-1$ is the subcarrier index, and $T$ is the OFDM symbol duration.

The frequency domain signal is transformed from serial to parallel and passed through an inverse FFT block, IFFT, to create a time domain parallel sequence.   It is desirable that the sequence length, $N$, be a power of $2$ for an efficient implementation of the FFT algorithm.  OFDM symbols representing the preamble and SFD are appended~\cite{homeplug,CKYK2010,HDW2004}.  Notice that after QAM/OFDM modulation, the generated time domain signal is clearly a Fourier series.  It is a sum of pure sinewaves multiplied by a gate function of duration $T$.  The time-domain signal of the combined subcarriers comprises one OFDM symbol.   A time domain burst carries $L$ OFDM symbols.  Thus, the $l$-th OFDM symbol, $s_l(t)$, can be written as a windowed sum of subcarriers,

{\small
\begin{equation}
  s_l(t)= \left\{
  \begin{array}{cc}
    Re\left\{{1\over T}\sum d_{k,l} e^{j2\pi\left(f_c - {k\over T}\right)(t-lt_s)}\right\},& lt_s<t<lt_s+T\\
    0,& \mbox{otherwise},
    \end{array} \right.
\end{equation}
}
  
Not all subcarriers, as defined by the IFFT block size, are used for data transfer.  Some subcarriers are pilot carriers, which transport known symbols, named pilot symbols, and are used for channel estimation and equalization.  Besides, some subcarriers are excluded due to poor channel quality for the particular subcarrier or to avoid interference with other services, such as amateur bands~\cite{homeplug,CKYK2010}.

Next, without loss of generality, single symbol transmission is considered.  Thus, for the $l= 1$ symbol, the transmitted signal over the channel can modeled as a discrete time Fourier series,

\begin{equation}
\mbox{IFFT:} \hskip0.5in b_1[n]= {1\over N}\sum_{k=0}^{N-1} d_{k,1} e^{j2\pi kn/N}.
\end{equation}

To avoid intersymbol interference, a cyclic prefix, CP, is appended to each OFDM symbol, i.e., the last $p$ samples of the inverse-Fourier-transformed sequence are repeated and appended to the sequence.  Notice that, CP is added in time domain.  The cyclic prefix works as guard band or symbol separation.   Another possibilities are cyclic suffix and zero padding~\cite{CKYK2010}.    With the introduction of CP, symbol duration is increased, $T'= T+T_G$.  The duration of CP is selected to be larger than channel delay, $T_G> \Delta$.  

\begin{equation}
  b_1[n] \Rightarrow \tilde{b}_1[n]
\end{equation}
in which, $\tilde{b}$ is the sequence with cyclic prefix.

Finally, the encoded parallel sequence is converted into an analog serial signal with a DAC and  the analog front-end generates the time domain OFDM burst, which is injected into the channel~\cite{HDW2004}. The real part of the time-domain signal is the in-phase carrier signal and the imaginary part is the out-of-phase carrier signal, as in Equation~\ref{eqno:04}.  One can calculate the power line raw PHY rate for 2$^n$-QAM/OFDM, as follows,

\begin{equation}
  \mbox{PHY rate}=  {\mbox{\# OFDM subcarriers}\times n\times \mbox{CR}\over (T+T_G)/2}
  \label{eq:phyrate}
\end{equation}
in which, $\mbox{\# OFDM subcarriers}$ is the number of available OFDM subcarriers, $n$ is the QAM modulation order, CR is the code rate, and $T_G$ is the guard interval.

\subsection{Channel transmission}

The channel modifies the signal burst, adding noise or other types of interferences, such as messages from distant nodes.   There are many models for noise and interference in the channel.  For white noise, Gaussian noise model is used, and for multipath interference, Rayleigh fading channel model is used.  The generalized Gaussian function is useful to represent different types of noise sources~\cite{SYA2013,HS2016}. 

\begin{equation}
  p_N(u|\mu, \sigma, \alpha)= {\alpha\Lambda\over 2\Gamma(1/\alpha)}e^{-\Lambda^\alpha|u-\mu|^\alpha}
\end{equation}
in which, $\alpha= 2$ for Gaussian noise, $\alpha= 0$ for impulsive noise.

\begin{equation}
\Lambda= {\Lambda_0\over\sigma}= \sqrt{\Gamma(3/\alpha)\over \sigma^2\Gamma(1/\alpha)}
\end{equation}
in which, for Gaussian noise, $\alpha= 2\Rightarrow \Lambda_0= 1/\sqrt{2}$.

Should noise interfere equally in all symbols, variance is given by

\begin{equation}
\sigma^2= \langle N^2\rangle - \mu^2= {{\cal N}_o\over 2}.
\end{equation}

The  symbol error rate, $\mbox{SER}$, is the probability of error, $\mbox{SER}= p_s$.  It is obtained from the generalized Q-function~\cite{SYA2013,HS2016}.

\begin{equation}
  Q_\alpha(x)= {\alpha \Lambda_0\over 2\Gamma(1/\alpha)}\int_x^\infty e^{-\Lambda_0^\alpha|u|^\alpha} du
\end{equation}

For Gaussian noise, the Q-function can be written as the complementary error function.

\begin{equation}
Q_2(x)= {1\over \sqrt{2\pi}}\int_x^{\infty}e^{-u^2/2}du= {1\over 2}\mbox{erfc}\left({x\over \sqrt{2}} \right)
\end{equation}

For M-QAM, the probability of error or symbol error rate~\cite{SYA2013,HS2016},

\begin{eqnarray}
  p_s({\cal E}_S)&=& 4\left(1-{1\over\sqrt{M}}\right)Q_2\left(\sqrt{3{\cal E}_S\over {\cal N}_o(M-1)}\right)\nonumber\\
  &&-4\left(1-{1\over\sqrt{M}}\right)^2Q_2^2\left(\sqrt{3{\cal E}_S\over {\cal N}_o(M-1)}\right)
  \label{eq:errorprob}
\end{eqnarray}
in which, ${\cal E}_S$ is the energy of the transmitted signal and ${\cal N}_o$ is the noise energy.

\begin{figure} [t] 
\centering
 \includegraphics[width=0.51\linewidth]{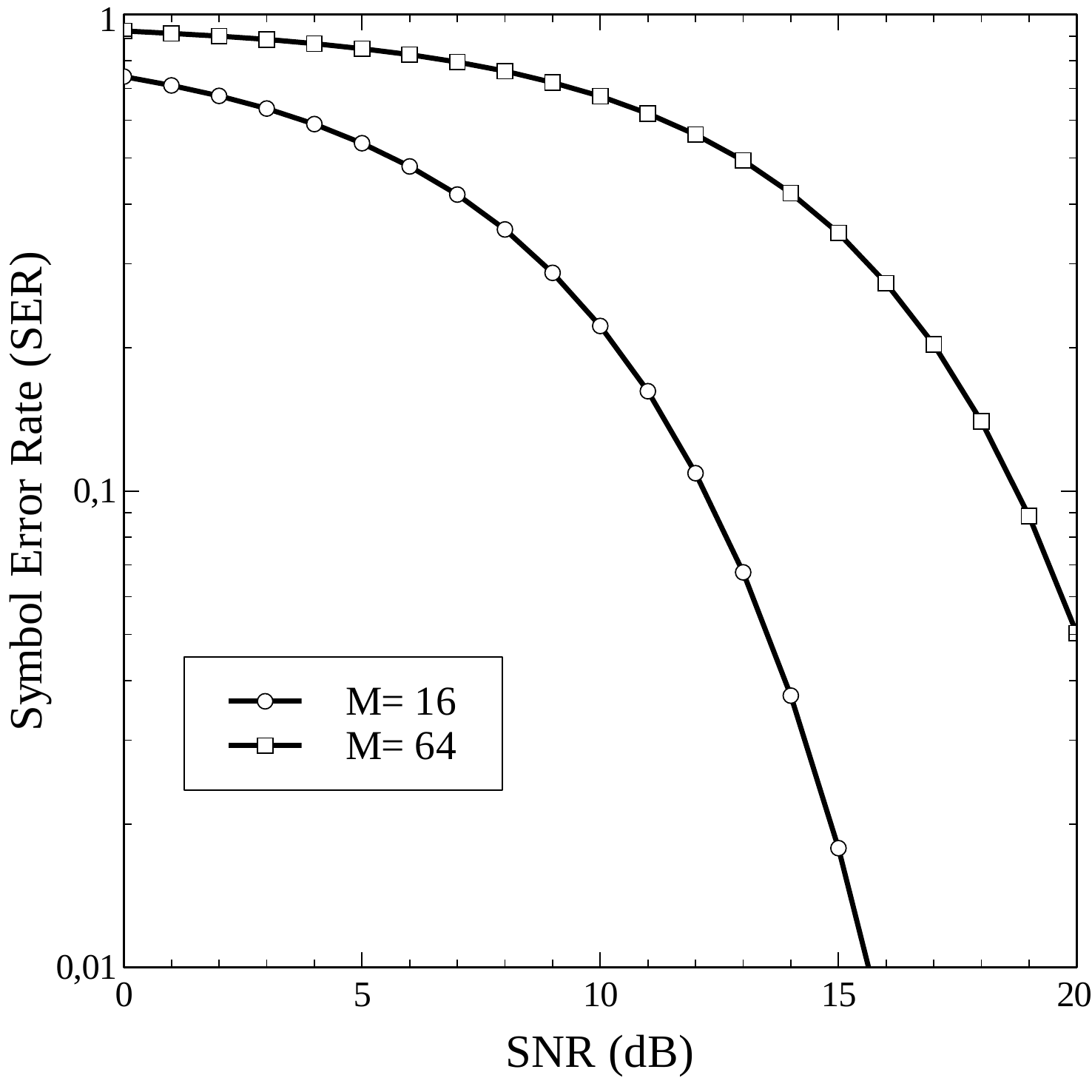} 
\caption{Symbol error rate for M=16 (2$^4$-QAM) and M=64 (2$^6$-QAM).}
\label{fig:SER}
\end{figure}

The $\mbox{SER}$, as a function of the signal to noise ratio in dB, $\mbox{SNR(dB)}= 10\log({\cal E}_S/{\cal N}_o)$ is presented in Figure~\ref{fig:SER}. Increasing the number of constellation points will increase data rate, but also increases the error rate.

 At receiver end, for a linear channel, the detected signal is the convolution of the original transmitted signal with the channel transfer function plus noise and interference from many sources. 

\begin{equation}
\tilde{R}_1[n]=  \sum_{m=0}^{N-1} h[m]\tilde{b}_1[n-\Delta-m] + \delta[n]
\end{equation}
in which, $h$ is the channel transfer function, $\Delta$ is the channel delay, and $\delta$ is the fluctuation due to channel noise or interference.

\subsection{Receiver}

\begin{figure} [t] 
\centering
 \includegraphics[width=0.51\linewidth]{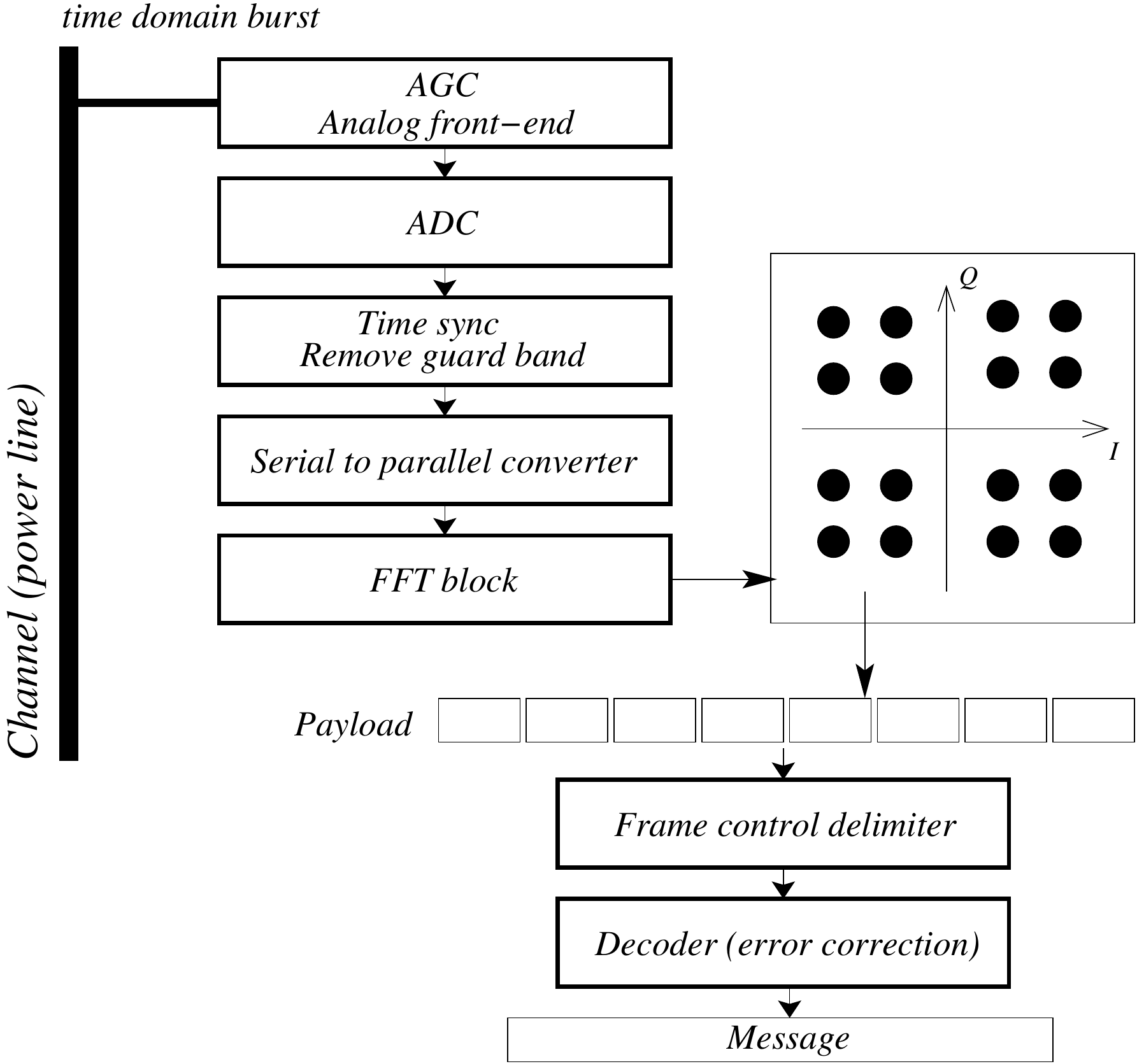} 
\caption{Block diagram of the QAM/OFDM receiver.}
\label{fig:OFDMreceiver}
\end{figure}

The receiver is always sampling the channel. The incoming signal, $\tilde{R}(t)$, is critically sampled at $N/T$, in which $N$ is the size of the Fourier block and $T$ is the symbol duration without the cyclic prefix.  Thus, converting back from analog to digital.   To detect channel activity the receiver performs self-correlation of collected samples for preamble detection~\cite{CKYK2010}. Preamble and SFD are known symbol sequences.  Preamble is used for carrier sensing, automatic gain control, clock frequency and offset estimation~\cite{homeplug,CKYK2010,SF2008}. If carrier is detected, the receiver carries out the needed tasks for synchronization.  It performs the opposite sequence of steps carried out at the transmitter to decode the message.  The simplified receiver block diagram is presented in Figure~\ref{fig:OFDMreceiver}.  Frame starts as self-correlation hits its maximum,

\begin{equation}
  \rho(k)= \sum_{n=k}^{k+N'-1}\tilde{R}_1[n]\tilde{R}_1[n-k]
\end{equation}
in which, $N'$ is the window size.

It can sample at full rate or at a reduced rate in case energy saving is required.  Preamble could be detected at lower sampling rate (downclocked rate) and switched at SFD to full clock rate to receive the packet, known as sampling rate invariant detection, SRID~\cite{ZS2012}.   If data transmission can be considered sparse, ADC and FFT/IFFT can further be optimized to reduce chip area~\cite{NP2020}. 

$N'$ data points, $R[n]$, are collected.  The samples corresponding to the cyclic prefix are discarded, $\tilde{R}[n] \Rightarrow R[n]$, and the resulting sequence is fed into the FFT block to obtain $D[k]$.

\begin{equation}
\mbox{FFT:}  \hskip0.5in   D[k]= \sum_{n=0}^{N-1} R[n] e^{-j2\pi kn/N}
\end{equation}

Next, the QAM decoder recovers the closest symbol.

\begin{equation}
D[k]\Rightarrow d_k
\end{equation}

The received signal may contain errors.  Forward error correction is carried out to remove white noise or burst noise induced errors.  At the receiver end, after demodulation and decryption, a cyclic redundancy check is carried out for data integrity.   If message is not for the node, it is discarded.

A difficulty with Fourier transformed signals is that time-bounded signal yields an infinite bandwidth frequency representation.  A finite bandwidth signal yields an infinitely long signal in the time domain.  A possible solution for this problem is to use the Wavelet transform~\cite{GKK2008,HMO2007}.  A practical implementation of this approach is HD-PLC developed by Panasonic, and later included in the IEEE 1901 standard, as IEEE 1901.1~\cite{IEEE1901}.

\bibliographystyle{model1-num-names}
\bibliography{<your-bib-database>}

\vfil

\end{document}